# Scanning Gate Spectroscopy on Nanoclusters

L. Gurevich[a], L. Canali, and L.P. Kouwenhoven

*Department of Applied Physics and DIMES, Delft University of Technology, P.O.Box 5046, 2600 GA Delft, the Netherlands*

A gated probe for scanning tunnelling microscopy (STM) has been developed.[b] The probe extends normal STM operations by means of an additional electrode fabricated next to the tunnelling tip. The extra electrode does not make contact with the sample and can be used as a gate. We report on the recipe used for fabricating the tunnelling tip and the gate electrode on a silicon nitride cantilever. We demonstrate the functioning of the scanning gate probes by performing single-electron tunnelling spectroscopy on 20-nm gold clusters for different gate voltages.

The constant pace in miniaturisation of electronic chips is thought[1] to eventually lead to circuit elements consisting of just a few molecules. For this reason a great effort is currently addressed to studying the transport properties of nanostructures such as quantum dots, fullerenes, nanotubes, metal nanoclusters and macromolecules. For such applications it is important to characterise the electronic energy spectrum of the nanostructures for different sizes and arrangements. At present, the interfacing of small objects with macroscopic electrical contacts presents the main experimental challenge, which has been successfully addressed only in some particular cases[2]. Conversely, scanning probe techniques, and scanning tunnelling microscopy (STM) in particular, do not suffer from interfacing problems. Typically the STM is used to image and locate nanostructures deposited on a conducting substrate. Afterwards the sharp STM tip can be selectively positioned on top of specific structures to acquire their spectroscopic characteristics[3]. This type of measurement can give valuable information on the electronic states of the molecule. A clearer interpretation, however, becomes possible when one is able to control and change the electronic states with the use of a gate electrode (i.e. perform three-terminal measurements). The gate can also be used to avoid the extra level smearing caused by high bias voltages and currents, by shifting the electronic levels in the nanostructure towards the Fermi energy of the leads[4]. Three-terminal electrical measurements have already been proven very fruitful in the field of SET transistors and semiconductor quantum dots[5].

Figure 1a illustrates the scheme of a typical STM measurement on a metallic cluster (or more generally on a nanostructure): electrons can tunnel from the STM tip onto the cluster and then across an insulating tunnel barrier into the conducting substrate. Inside the cluster the extra electron can occupy the ground state or, when a voltage bias is applied, one of the excited states. No current can flow if the tunneling electrons have energy less than the charging energy of the cluster ($E_c = e^2/C$, where $C$ is the capacitance of the cluster). $E_c$ can be large for small clusters. The charging energy strongly regulates transport, causing electrons to tunnel one-by-one, which is known as single electron tunneling[5].

So far, the gate electrode has been absent in all STM studies. In this Letter we present the design and operation of a scanning probe, which extends normal STM capabilities by means of an extra gate electrode with which we are able to acquire three-terminal spectroscopic characteristics of nanostructures deposited on a conducting surface.

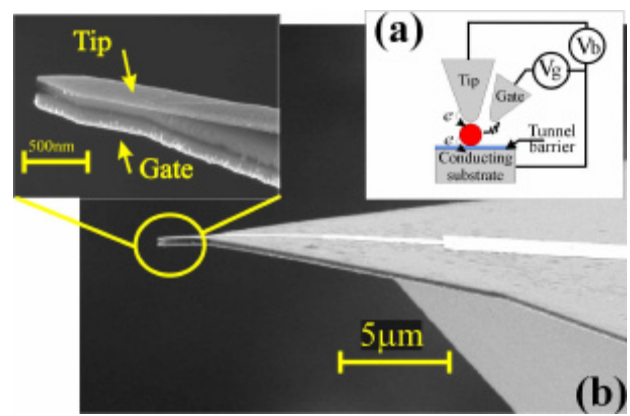

FIG.1. (a) Scheme of the scanning gate probe. The STM tip is positioned above a metallic cluster (represented as a sphere). An electron coming from the tip will first tunnel onto the cluster and then tunnel into the conducting substrate. A gate electrode can be used to change the offset charge of the cluster, thus forming a system equivalent to a single-electron transistor. (b) SEM image of a scanning gate probe. Metal electrodes are deposited on the two sides of a $SiN_x$-$SiO_2$-$SiN_x$ cantilever. The separation between the electrodes is about of 200 nm and the typical radius of curvature of the sharp end of the tip is approximately 30 nm.

We microfabricate the scanning gate probes using a five-step process involving e-beam and optical lithography. We start with a Si (100) wafer covered on both sides with a multilayer of $SiN_x$-$SiO_2$-$SiN_x$, 100 nm per layer, deposited by low-pressure chemical vapour deposition (see inset to Fig. 2a). First, we pattern the backside of the wafer by optical lithography and then we etch through the multilayer by reactive ion etching using $CHF_3$. Afterwards we etch the silicon wafer through its entire thickness in a 30%-KOH aqueous solution with addition of isopropanol at 80° C. The solution does not etch the top $SiN_x$-$SiO_2$-$SiN_x$ multilayer, so that a thin membrane is formed on the topside of the wafer (Fig. 2b). Next we fabricate the Pt tip electrode by e-beam lithography, evaporation and lift-off. The tip is aligned to the sharp end of the silicon chip (Fig. 2c). Thereafter an Al-mask is fabricated on top of the membrane and aligned to the tip electrode, by means of e-beam lithography, evaporation and lift-off. We use the mask to protect the tip while we etch through the multilayer by anisotropic plasma etching with $CHF_3$ and form a sharp, freestanding cantilever (Fig. 2d). Subsequently we underetch the oxide part of the $SiN_x$-$SiO_2$-$SiN_x$ multilayer by dipping the wafer in BHF solution. This is done to prevent unwanted electrical contacts between the final top and bottom electrodes (inset to Fig. 2e). Last we evaporate the platinum gate electrode on the backside of the chip (Fig. 2e). After fabrication the scanning gate probes can be mechanically detached from the wafer. The probes consist of two closely separated metal electrodes (with a distance of about 200 nm) extending onto the sharp ends of a silicon nitride cantilever (Fig.1b). The whole chip has size similar to commercial tips used for atomic force microscopy (AFM) ~ 2 x 4 x 0.5 $mm^3$. They are easy to handle and can be installed in standard STM heads.

We demonstrate the functioning of the scanning gate probes by performing spectroscopic measurements on a test sample consisting of 20-nm gold clusters deposited on a flat platinum film. The sample is prepared taking into account the following constraints. The clusters have to be fixed on the surface during scanning and spectroscopic measurements. There must be an insulating layer between the clusters and the metal substrate to provide a tunnel-barrier. The insulator must have a smooth *I-V* characteristic in the bias window of interest for spectroscopy on nanoclusters (about 1 V). Finally, the clusters must be well separated from each other, otherwise their capacitive coupling to the gate would be reduced and some parasitic potential jumps caused by moving electrons in the neighbouring clusters might occur. These requirements are fulfilled by depositing 20-nm gold clusters[6] from a charge-stabilised aqueous solution onto a Pt substrate covered by a self-assembled monolayer of cysteamine. This monolayer both bonds the clusters to the substrate and forms the tunnel barrier between the clusters and the Pt film (Fig. 3a). After deposition the samples are imaged by tapping mode AFM to find the optimal concentration of clusters on the substrate, as shown in Fig. 3b.

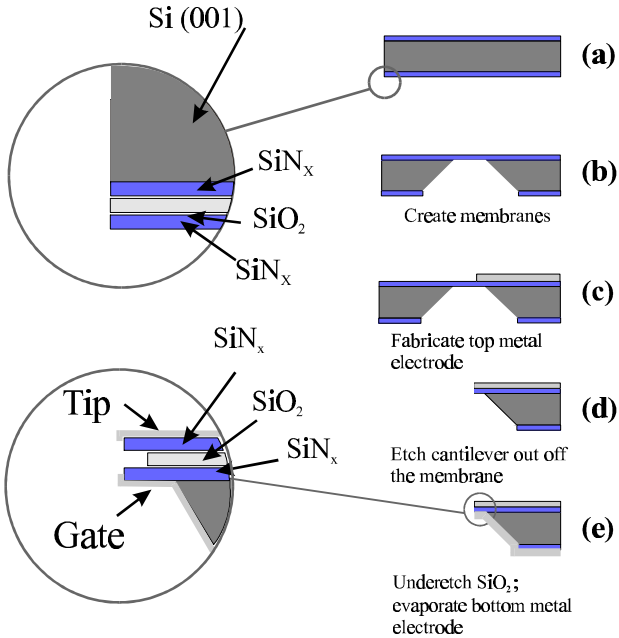

FIG.2. Fabrication steps for the scanning gate probes. (a) The Si (100) wafer is covered on both sides with a multilayer of $SiN_x$-$SiO_2$-$SiN_x$. (b) A thin membrane is formed on the topside of the wafer. (c) The metallic tip is fabricated on the sharp end of the silicon chip. (d) The sharp freestanding cantilever is etched out of the wafer. (e) The gate electrode is evaporated.

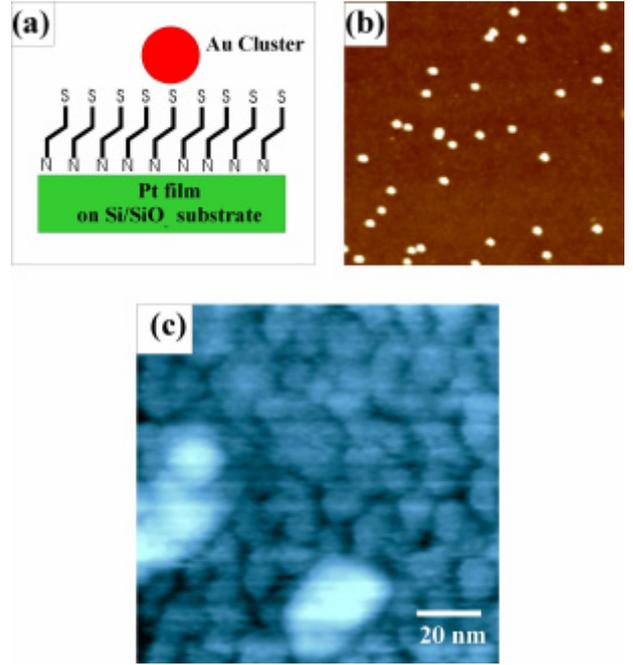

FIG.3. (a) Schematic diagram of the sample layout. The Au cluster is chemically bonded to the surface by a cysteamine ($C_2H_7NS$) self-assembled monolayer. The monolayer also acts as a tunnel barrier. (b) Tapping mode AFM image of the sample; 20-nm Au clusters are visible as bright spots. The scan field is 1x1 $\mu m^2$. (c) STM topography of the sample imaged with a scanning gate probe at 4.2 K. A few 20-nm Au clusters are visible in the scan area.

The probe and sample are mounted on a home-built STM (design similar to Ref. 7), which is attached to the cold finger of a commercial $^3$He fridge. Figure 3c shows a topographic image obtained on such a sample with a scanning gate probe at 4.2 K. STM topographies are acquired with a relatively high tunnel resistance (and thus large tip-sample distance) of the order of 10 G$\Omega$, to prevent the clusters from moving on the surface. From the topographic images we find the position of the gold clusters so that we can position the scanning gate on top of one of them and measure its *I-V* characteristic.

Figure 4 (top graph) shows two such *I-V*s obtained for different gate voltages, corresponding to an offset charge on the cluster $Q_0 = 0$ and e/2. Coulomb charging steps are clearly visible, with a step size of about 15 meV (as expected for 20-nm metallic clusters). We acquired a set of 50 *I-V*s for different gate voltages varying from -30 V to +30 V using the STM program. Before each *I-V* curve, the STM tip is stabilized by the feedback loop with bias voltage $V_b$ = 100 mV, tunnel current $I_T$ = 50 pA and gate voltage $V_g$ = 0. Then the feedback loop is turned off and the gate voltage is swept to the intended value. Afterwards we acquire the *I-V* data points, sweep the gate back to zero and turn the feedback on again to stabilize the STM tip.

The measurements are shown as a 3d colour-plot map in Fig. 4. Such a figure is called the stability diagram of a single-electron transistor[8]. The dark area at the centre of the image corresponds to the Coulomb blockade of transport: the current through the cluster is blocked because the electrons in the leads do not have enough energy to tunnel onto the cluster (their energy is lower than the charging energy of the cluster). Areas represented with different colours correspond to processes where one or more electrons can tunnel simultaneously through. The discrete electronic spectrum of the cluster could not be resolved, as expected, since the mean electron-level separation for a 20-nm cluster can be estimated to be less than 1 K. Using the *orthodox theory* of single electron tunnelling[8] we were able to calculate the values for the capacitances and resistances of the tunnel barriers and for the gate capacitance (see the caption of Fig. 4).

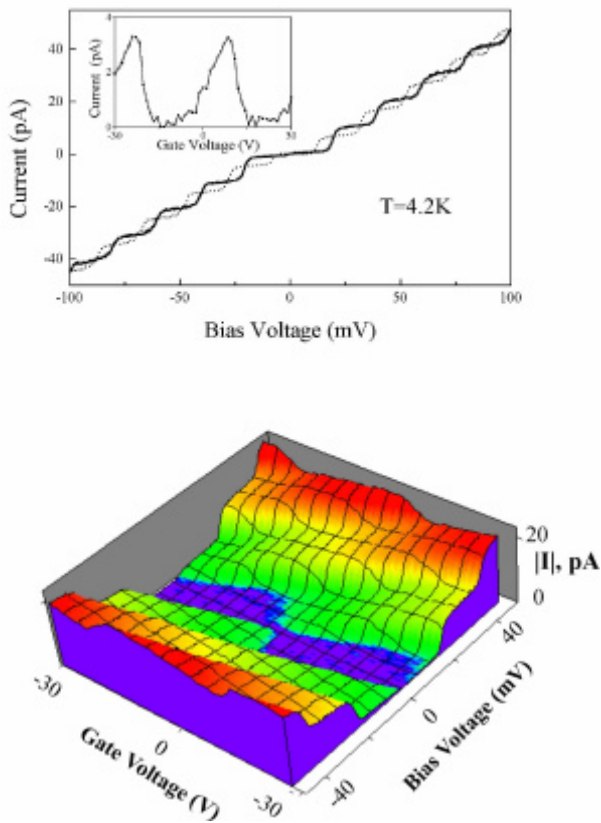

FIG. 4. Scanning gate spectroscopy of a 20-nm Au cluster at 4.2 K. Top graph: *I-V* curves taken with the scanning gate tip positioned on top of a cluster for two different gate voltages $V_g$ = -2.4 and 15.6 V, corresponding to an induced charge on the cluster $Q_0$ = 0 and $e/2$ respectively. Inset: Current versus gate voltage characteristic with bias voltage $V_b$ = 8.4 mV. Lower image: Complete set of I-V curves represented as a greyscale map. The barrier capacitances and resistances estimated from the data are: $C_1$ = 8×10$^{-18}$ F, $C_2$ = 1×10$^{-18}$ F, $C_g$ = 5×10$^{-21}$ F, $R_1$ = 2 GΩ, $R_2$ = 50 MΩ, where $C_1$ and $R_1$ refer to the tip-cluster barrier, $C_2$ and $R_2$ refer to the cluster-substrate barrier and $C_g$ is the gate-cluster capacitance.

In conclusion, we have developed a scanning gate probe for studying nanostructures such as molecules and metallic clusters. The probe can image and perform three-terminal spectroscopy on nanostructures deposited on a conducting substrate. We report on the recipe used to microfabricate the scanning gate probes on a silicon nitride cantilever and on the test measurements performed on 20-nm Au clusters, where we could demonstrate the correct functioning of the probe. The scanning gates can be applied to the study of a large spectrum of nanostructures both at room and low temperature. It is also conceivable to improve the fabrication recipe as to reduce the tip-gate distance and thus increase the gate-sample capacitance by up to an order of magnitude. We acknowledge fruitful discussions and experimental help of Alexey Bezryadin, Yann Kervennic, Gesha Kravchenko, Hans Mooij, Danny Porath, Arnold van Run, Anja Suurling, Jeroen Wildöer, and Hugo de Wit. We would like to thank Günter Schmid, Monika Bäumle and Paul Weiss for enlightening advice on the chemistry of clusters and self-assembled monolayers. This work was supported by the Dutch Stichting voor Fundamenteel Onderzoek der Materie (FOM).